\documentclass[aip,jap,reprint,groupedaddress]{revtex4-1}
\draft 
\usepackage{amsmath}
\usepackage{amssymb}
\usepackage{graphicx}
\usepackage{tikz}
\usepackage{multirow}
\begin{document}
\title{A 490~GHz planar circuit balanced
Nb-Al$_\mathbf{2}$O$_{\mathbf{3}}$-Nb 
quasiparticle mixer for radio astronomy: 
Application to quantitative local oscillator noise determination}
\author{M.~P.~Westig}
\email[]{westig@ph1.uni-koeln.de}
\affiliation{K\"olner Observatorium f\"ur Submillimeter Astronomie 
(KOSMA), I.~Physikalisches Institut, 
Universit\"at zu K\"oln, 
D-50937 K\"oln, Germany}
\author{M.~Justen}
\affiliation{K\"olner Observatorium f\"ur Submillimeter Astronomie 
(KOSMA), I.~Physikalisches Institut, 
Universit\"at zu K\"oln, 
D-50937 K\"oln, Germany}
\author{K.~Jacobs}
\affiliation{K\"olner Observatorium f\"ur Submillimeter Astronomie 
(KOSMA), I.~Physikalisches Institut, 
Universit\"at zu K\"oln, 
D-50937 K\"oln, Germany}
\author{J.~Stutzki}
\affiliation{K\"olner Observatorium f\"ur Submillimeter Astronomie 
(KOSMA), I.~Physikalisches Institut, 
Universit\"at zu K\"oln, 
D-50937 K\"oln, Germany}
\author{M.~Schultz}
\affiliation{K\"olner Observatorium f\"ur Submillimeter Astronomie 
(KOSMA), I.~Physikalisches Institut, 
Universit\"at zu K\"oln, 
D-50937 K\"oln, Germany}
\author{F.~Schomacker}
\affiliation{K\"olner Observatorium f\"ur Submillimeter Astronomie 
(KOSMA), I.~Physikalisches Institut, 
Universit\"at zu K\"oln, 
D-50937 K\"oln, Germany}
\author{C.~E.~Honingh}
\affiliation{K\"olner Observatorium f\"ur Submillimeter Astronomie 
(KOSMA), I.~Physikalisches Institut, 
Universit\"at zu K\"oln, 
D-50937 K\"oln, Germany}
\date{\today}
\begin{abstract}
This article presents a heterodyne experiment which uses a 
380-520~GHz planar circuit balanced Nb-$\mathrm{Al_2O_3}$-Nb 
superconductor-insulator-superconductor (SIS) quasiparticle 
mixer with 4-8~GHz instantaneous intermediate frequency (IF) 
bandwidth to quantitatively determine local oscillator (LO) 
noise. A balanced mixer is a unique tool to separate noise at the 
mixer's LO port from other noise sources. This is not possible in 
single-ended mixers. The antisymmetric IV characteristic of a SIS
mixer further helps to simplify the measurements.
The double-sideband receiver sensitivity of the balanced mixer 
is 2-4 times the quantum noise limit $h\nu/k_B$ over the 
measured frequencies with a maximum LO noise rejection of 
15~dB. This work presents independent measurements with three
different LO sources that produce the reference frequency but also 
an amount of near-carrier noise power which is quantified in the 
experiment as a function of the LO and IF frequency in terms of an
equivalent noise temperature $T_{LO}$.
Two types of LO sources are used: a synthesizer driven 
amplifier/multiplier chain and a Gunn oscillator driven 
multiplier chain. With the first type of LO we find different 
near-carrier noise contributions using two different power 
pre-amplifiers of the LO system. For one of the two 
power pre-amplifiers we measure $T_{LO} =30\pm4$~K 
at the LO frequency 380~GHz and $T_{LO} = 38\pm10$~K at 420~GHz. 
At the frequency band center 465~GHz of the Gunn driven LO 
we measure a comparable value of $T_{LO} = 32 \pm 6$~K. 
For the second power pre-amplifier a significant 
higher $T_{LO}$ value of the synthesizer driven LO is found 
which is up to six times larger compared with the best values 
found for the Gunn driven LO. In a second experiment we use 
only one of two SIS mixers of the balanced mixer chip in order to
verify the influence of near-carrier LO noise power on a single-ended 
heterodyne mixer measurement. We find an IF frequency dependence of 
near-carrier LO noise power. The frequency-resolved IF noise 
temperature slope is flat or slightly negative for the 
single-ended mixer. This is in contrast to the IF slope 
of the balanced mixer itself which is positive due to the 
expected IF roll-off of the mixer. This indicates a higher noise 
level closer to the LO's carrier frequency. Our findings imply that 
near-carrier LO noise has the largest impact on the sensitivity 
of a receiver system which uses mixers with a low IF band, 
for example superconducting hot-electron bolometer (HEB) mixers. 
\end{abstract}
\maketitle 
\section{Introduction}
\label{sec:001}
State-of-the-art low-noise heterodyne receivers are
a key technology in radio astronomy and are 
needed to conduct high-resolution spectroscopy 
($\nu/\delta \nu \geq 10^{6}$) in the 
0.3 - 5~terahertz (THz) frequency band. 
Here, among many fundamental astronomical objects, 
for example, star-forming regions 
have very rich atomic and molecular 
spectra\cite{blain2002}. The highest possible 
receiver sensitivity is needed because the signals
which contain the astronomical information are usually very
weak. For spectroscopy in the frequency band 0.3 - 1.2~THz, 
heterodyne receivers employ superconductor-insulator-superconductor (SIS) 
detectors as frequency mixing devices. They are 
operational up to frequencies of $4\Delta/h$, 
where $\Delta$ is the superconducting gap energy of the detector 
electrodes. For frequencies larger than $4\Delta/h$ where SIS 
devices do not work anymore, superconducting hot-electron 
bolometers (HEB) are used~\cite{zmuidzinas2004}. 
In a THz heterodyne receiver a weak signal with
frequency $\nu_S$ is detected by multiplying it with 
a strong local oscillator (LO) reference signal with 
frequency $\nu_{LO}$ in the SIS or 
HEB device where the intermediate frequency (IF) 
$\nu_{IF} = \vert{\nu_{S} - \nu_{LO}}\vert$ 
is produced. Generally, $\nu_{IF}$ ranges from 
0 - 4~GHz for an HEB device or from 
4 - 12~GHz for an SIS device. Heterodyne detection 
preserves amplitude and phase of the incoming signal. 
Therefore, Heisenberg's uncertainty principle imposes a 
fundamental limit\cite{caves1982} on the sensitivity usually 
expressed by a quantum noise temperature 
$T_{qn} = h\nu_{S}/k_B$ for double-sideband operation, 
$h$ and $k_B$ are the Planck and Boltzmann constants. 

For heterodyne receivers, presently two established LO 
technologies are used. Gunn oscillator driven multiplier 
chains\cite{gunn1963} for frequencies up to approximately 
0.8~THz and synthesizer driven amplifier/multiplier chains 
that are easier to operate and which are 
used for almost the whole THz frequency range, nowadays even 
exceeding 2~THz\cite{vdi2012,pearson2011}. A Gunn driven 
LO consists of a III-V semiconductor (usually GaAs or InP) 
embedded in a mechanically tunable high-Q waveguide resonator 
which produces amplitude stable signals 
with high spectral purity and a very low amount
of near-carrier noise power. A cascade of Schottky 
diode frequency multipliers is used 
to produce the output frequency of the Gunn driven LO.

However, receivers in modern radio-astronomy 
experiments increasingly use synthesizer driven 
LO's due to their easy handling. First, the 
signal of the synthesizer is fed into a 
frequency multiplier, usually a doubler or 
a tripler. The output is connected to a power pre-amplifier 
which is connected to a cascade of frequency 
multipliers. With this technique frequencies 
up to approximately 2~THz can be produced.
Mainly stimulated by receiver developments
for ground-based and space observatories 
where strict performance 
specifications have to be fulfilled, it was found 
that synthesizer based LO sources have to be 
operated and designed following rules which are 
summarized in\cite{erickson2004,bryerton2007,
bryerton2008}. An important conclusion is that
near-carrier LO noise power in these devices 
can be minimized if the input power saturates the LO 
components, i.e.~the power pre-amplifier and the multiplier chain. 

In the near future quantum cascade lasers (QCLs\cite{faist1994}) 
might offer another LO technology solving the problem of low
output power for frequencies larger than 
1~THz,\cite{scalari2010,turcinkova2011} rendering the development of 
heterodyne cameras possible. Due to the laser operation
a very low noise is expected for a QCL.

The most sensitive heterodyne receivers operate 
at a few times the quantum noise limit. 
Each element of the receiver, usually consisting 
of quasioptics to focus the signal and the LO 
beam on the mixer, followed by a low-noise cryogenic IF 
amplifier,\cite{bardin2009,weinreb2009,
wadefalk2005} additional low-noise amplifiers at 
room temperature and a spectrometer adds a noise contribution 
to the total receiver noise temperature $T_{rec}$. This can 
be analyzed by decomposition according to the {\it Friis} 
formula~\cite{freeman1958}. 
In practice, $T_{rec}$ is measured with a standard $Y$-factor 
method\cite{rohlfs1999}. LO near-carrier noise power which 
is among other things caused by a fluctuating LO amplitude (AM noise),
is mixed down into the IF band, thus 
has a substantial impact on $T_{rec}$ and results in a 
decrease of the receiver's sensitivity.
However, in a heterodyne receiver with a 
single-ended mixer an unambiguous separation 
and quantification of the 
unknown amount of near-carrier LO noise power 
from other receiver noise contributions is not 
possible. Any noise from the LO sidebands is 
downconverted in the frequency 
mixing process and is at the end indistinguishable 
from the desired IF signal\cite{bryerton2007}. 
This shows that a single-ended mixer is very 
susceptible to this particular noise source. 

Balanced mixers are much less sensitive to near-carrier LO noise 
power and offer the possibility of a direct quantification of 
the noise produced by the complete LO system. With this mixer 
technology it is possible to separate the LO AM noise from other
noise sources of the heterodyne receiver. In the submm range this 
was not possible for a long time due to the complexity of 
building sensitive balanced mixers for these small wavelengths.

This paper is 
organized into four sections. Following the 
introduction, Sec.~\ref{sec2} describes the 
experimental setup and the noise measurement 
principle using a planar circuit balanced 
quasiparticle mixer device\cite{westig2011}.
Section~\ref{sec3} presents the experiment.
In the 380-520~GHz range we measure the near-carrier 
noise power produced by three different LO 
sources: a Gunn driven LO and a synthesizer driven 
LO with two different power pre-amplifiers. 
We show how to separate this unwanted noise contribution 
from the receiver system and quantify the 
equivalent near-carrier noise temperature $T_{LO}$ by 
using our balanced mixer device as a noise 
meter. The impact of near-carrier noise power on 
single-ended mixers is studied in a second
experiment and is used to verify our results. Here 
we observe that the IF spectra of the single-ended-
and the balanced mixer measurements differ significantly.
Sec.~\ref{sec4} discusses the results 
and presents our conclusions. Especially 
the conclusion with regard to the spectral characteristic
of the LO noise is relevant for THz HEB mixers that
generally operate with a low IF band and where 
the balanced technology is not yet broadly established.
In these mixers near-carrier noise power from the LO
can substantially degrade the receiver's 
sensitivity.
\section{\label{sec2}Description of the Experiment}
\subsection{\label{sec2A}Experimental setup}
\begin{figure*}[tb]
\includegraphics[width=0.85\textwidth]{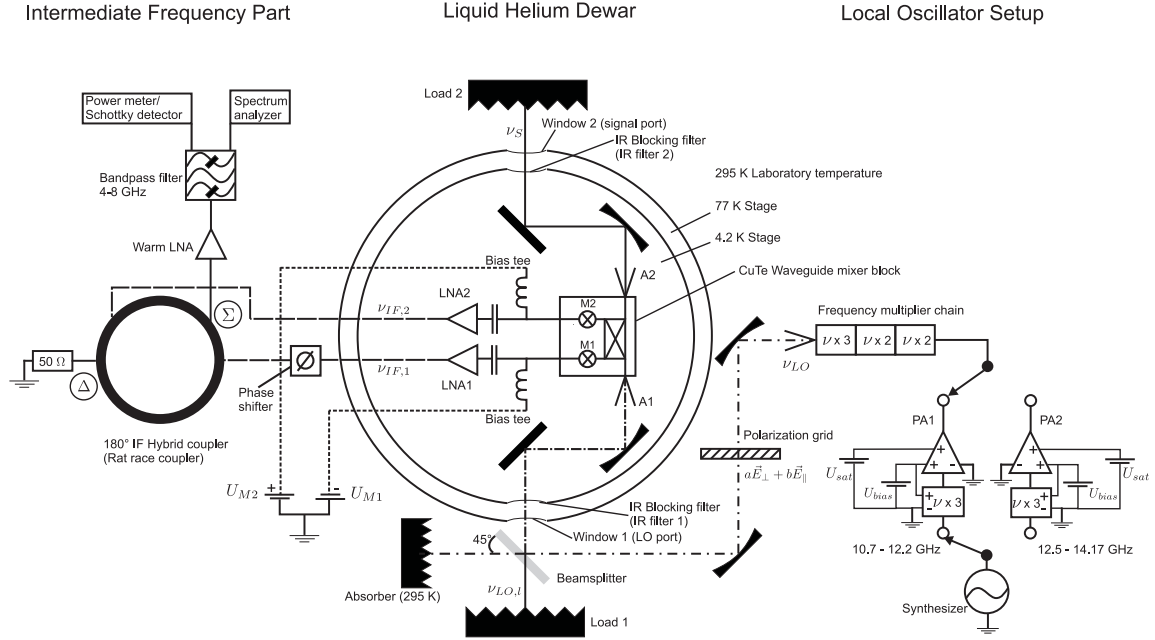}
\caption{\label{fig:01}Sketch of the experimental setup. 
The two SIS mixers (M1 and M2) of the 
balanced circuit are indicated with the 
$\bigotimes$-symbol and the crossed rectangle 
is the $90^{\circ}$ hybrid coupler
(compare also with Fig.~\ref{fig:03}).
The combination of the separately amplified IF 
signals of the balanced mixer ($\nu_{IF,1}$ 
and $\nu_{IF,2}$) is done at room temperature 
(295~K) with a commercial $180^\circ$ IF hybrid 
coupler. Two separate blackbody 
loads, load 1 (LO port) and load 2 (signal port), 
are used at a temperature of either 77~K or 295~K. Their 
noise power is used to measure $T_{rec}$ 
and near-carrier LO noise power simultaneously. 
The measurement is performed with a bias 
voltage sweep of either $U_{M1}$ or $U_{M2}$ while the 
other mixer bias voltage is kept 
constant within the first photon-assisted 
tunneling step of the SIS IV characteristic. The 
combined IF power is read-out at the $\Sigma$ output 
of the $180^{\circ}$ IF hybrid coupler. In the synthesizer
driven LO a switch is used to choose one of the two power 
pre-amplifiers (PA1 or PA2) for the measurement. 
In this experiment load 1 is used as a 77~K termination
of the LO port in order to determine the LO noise and is 
removed in an astronomical receiver. Windows, IR blocking filters and
beamsplitter each have a frequency dependent transmissivity which is
determined in Sec.~\ref{sec2B}.}
\end{figure*}
Fig.~\ref{fig:01} shows the experimental setup. 
The sketch of the LO shows the most important 
parts of this device, 
developed by Virginia Diodes Inc\cite{vdi2012}. 
We use a synthesizer\cite{synthesizer} 
with a measured phase noise of 115~dBc/Hz, 90~kHz away 
from the carrier frequency set to 11.806~GHz.
We could not detect an amplitude noise contribution 
from the synthesizer in a time domain measurement with a 
spectrum analyzer\cite{spectrumanalyzer}. The synthesizer 
is tuned either to a frequency between 
10.7 - 12.2~GHz or to a frequency between 12.5 - 14.17~GHz 
and is connected to one of two actively biased 
frequency triplers. No frequency standard was
used during our measurements. Each of the
two triplers is connected to one of two power 
pre-amplifiers\cite{spaceklabs} PA1 and PA2. 
The power pre-amplifiers are biased with 
$U_{bias} =$ 8 - 12~V. A second bias voltage
$U_{sat} =$ 0 - 5~V feeds an electronic attenuation
circuit. For $U_{sat} = 0$~V the maximum output power 
and for $U_{sat} = 5$~V the maximum attenuation of 
PA1 and PA2 is chosen. A feedback loop (not shown in 
Fig.~\ref{fig:01}) measures the power delivered
to the first module of the frequency multiplier 
chain, sets the output power correctly and protects
it from an excess of input power. 
The frequency multiplier chain consists of
three modules: two frequency doublers and a frequency 
tripler. For measurements using a Gunn driven LO, we 
replace the synthesizer driven LO. A rotatable polarization grid 
is used to attenuate the LO power independently 
from applying the voltage $U_{sat}$. 

The mixer is only 
sensitive to the $\vec{E}_{\perp}$ component of the LO's 
electric field which points perpendicularly out of the 
paper plane in Fig.~\ref{fig:01}. A $21~\mu$m thick 
Mylar foil is used as a beamsplitter to 
feed the LO signal into the dewar and provides the possibility 
to terminate window 1 (LO port) with a thermal load of varying
temperature (load 1 in Fig.~\ref{fig:01}). 
Window 1 consists of a 426~$\mu$m thick slab of Teflon material. 
The IR blocking filter on the 77~K radiation 
shield behind window 1 is a slab of 267~$\mu$m thick high-density 
polyethylene (HDPE). Window 2 (signal port) is made of a 
482~$\mu$m thick slab of Teflon and behind this window 
is a 261~$\mu$m thick HDPE IR blocking filter, also on the
77~K radiation shield. The balanced mixer chip is assembled
inside a gold plated tellurium copper split-block 
full-height waveguide mixer block. This mixer block is fixed in 
a cold optics assembly on the liquid helium (LHe) dewar 
4.2~K cold stage. Load 1 in front of window 1 is used to 
terminate the LO port with a 77~K load throughout the experiment. 
Noise power from this load is superimposed with the LO signal.
The thermal load in front of window 2 (load 2) is either 
cooled to 77~K or replaced by a 295~K load and serves as a 
calibration blackbody radiation source, representing the sky signal.
LO signal and noise power from load 1 are transmitted through
window 1 (LO port) while noise power from load 2 is transmitted through
window 2 (signal port). Both signals are
received by the balanced mixer via two waveguide horn antennas 
A1 and A2 (also referred to as "mixer input 
ports" in the text). The signals are coupled into 
the balanced mixer chip where they are superimposed, 
phase shifted and equally 
distributed (-3~dB) to the two individual SIS mixers 
M1 and M2. Here the IF signals 
$\nu_{IF,1}$ and $\nu_{IF,2}$ are produced and are
separately amplified by two MMIC WBA13 low noise amplifiers 
LNA1 and LNA2\cite{wadefalk2005}. Outside of the LHe 
dewar the two IF signals are combined by a $180^{\circ}$ 
IF hybrid coupler\cite{ratracecoupler}. A phase shifter 
which is connected to the IF output 
port of M1 is used to adjust the phase 
between the IF frequency paths of M1 
and M2 in order to achieve the best possible 
balanced performance. The combined IF signal is read-out at 
the $\Sigma$ output (sum port) of the $180^{\circ}$ 
IF hybrid coupler, and the 
$\Delta$ output (difference port) is terminated by a 
$50~\Omega$ load throughout the experiment.
We use a power meter and a power calibrated
Schottky detector, respectively, to measure the total
IF output power over 4-8~GHz as a function
of mixer bias voltage $U_{M1}$ or $U_{M2}$ 
while the other mixer is constant voltage
biased within the first photon-assisted 
tunneling step of the SIS IV curve. Therefore, in a single
bias voltage sweep of one of the two mixers, the balanced
mixer provides two different measurement results. At the 
$\Sigma$ output of the IF hybrid coupler for same bias polarity
of mixers M1 and M2, LO noise power and the noise power
coming from load 1 (LO port) is measured. $T_{rec}$ is measured
independently for opposite bias polarity of M1 and M2 at the same
port of the IF hybrid coupler. For further details
we refer to Sec.~\ref{sec2C}. A spectrum analyzer is used to measure the
frequency resolved IF power signal with fixed 
mixer bias voltage where now both mixers are voltage
biased within the first photon-assisted tunneling step. 
\subsection{\label{sec2B}Effective load temperature}
\begin{table}[tb]
\caption{\label{tab:01}Summary of the parameters used
to determine the transmissivity and emissivity of the windows, 
IR blocking filters and the beamsplitter in the measurement
bandwidth 380 - 520~GHz. The beamsplitter is rotated by
$45^{\circ}$ with respect to the LO beam axis. $d$ is the 
thickness and $n_d$ is the complex refractive index. 
The uncertainty (not shown in the table)
of the values below is considered in the 
calculation of the uncertainty of the effective temperatures, 
summarized in Table~\ref{tab:02}.}
\begin{ruledtabular}
\begin{tabular}{lccc}
&$d~[\mu\mathrm{m}]$& $n_d$ & Material\\
\hline
Window 1 & 426 & $1.46 - i~0.0018$ & Teflon\\
IR filter 1 & 267 & $1.32 - i~0.0005$ & HDPE\\
Beamsplitter & 21 & $1.80 - i~0.0320$ & Mylar\\
Window 2 & 482 & $1.52 - i~0.0020$ & Teflon\\
IR filter 2 & 261 & $1.22 - i~0.0005$ & HDPE\\
\end{tabular}
\end{ruledtabular}
\end{table}
To determine the effective load temperatures 
(Rayleigh-Jeans limit) at the mixer input ports 
A1 and A2, respectively, the transmissivity and emissivity of
the IR blocking filters and the vacuum windows in front of
both input ports and of the beamsplitter in front of A1 
have to be known. The IR blocking filters have a 
temperature of approximately $85$~K, measured in an earlier 
experiment with a temperature sensor. Window 1, window 2 
and the beamsplitter have a temperature of 295~K. 
Transmissivity is measured using a Fourier transform spectrometer 
(FTS) working in the frequency range 1 - 6~THz from which we obtain the 
frequency dependent gain $G(\nu) < 1$ of each element. 
Moreover, in our high-frequency measurement bandwidth the individual 
complex refractive index $n_d = n_d' - i n_d''$ of the IR blocking filters, windows 
and the beamsplitter is not known. Furthermore, polishing of the 
dielectric materials to the nominal size is expected to further modify the 
dielectric properties. Fourier transform spectrometry provides a straightforward 
determination method of the individual characteristics of each element.
Finally, the {\it Fresnel} theory is used to fit the experimental 
data and to extrapolate the transmissivity to our measurement 
bandwidth 380 - 520~GHz where the FTS detector is not 
sensitive enough for a direct measurement. For the 
beamsplitter the transmissivity is evaluated for an angle 
of $45^{\circ}$ relative to the LO beam axis. Free parameters of the 
fit are the thickness $d$ and the complex refractive index of the 
slab, the results are summarized in Table~\ref{tab:01}. 
The results of the fits are used further to calculate the effective load temperatures 
as a function of frequency.

\begin{figure}[tb]
\includegraphics[width=\columnwidth]{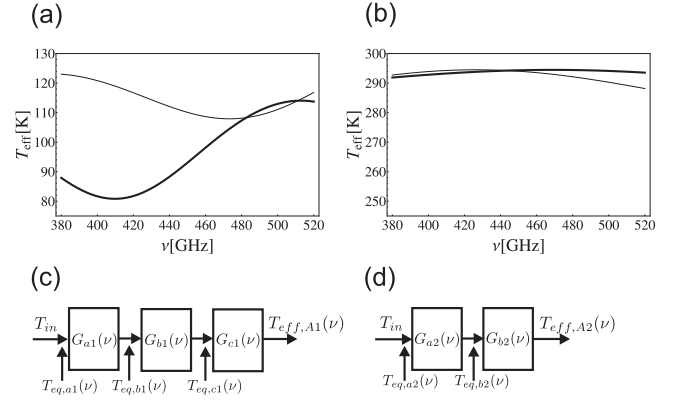}
\caption{\label{fig:02}
In (a) for $T_{in} = 77$~K and
in (b) for $T_{in} = 295$~K the effective load temperatures
are calculated for frequencies in the measurement bandwidth. 
The thin solid lines in (a) and (b) show the effective 
load temperature $T_{eff,A1}$ referred to mixer input port A1 
and the thick solid lines show $T_{eff,A2}$ referred to 
mixer input port A2. Equivalent circuit diagrams for
the window-IR filter-beamsplitter and window-IR filter 
cascade in front of mixer input ports A1 and A2 
are shown in (c) and (d).}
\end{figure}

A thermal load having a temperature $T_{in}$ of either
$77$~K or $295$~K in front of 
window 1 yields at the mixer input port A1 
an effective temperature of
\begin{equation}
\label{eq:01}
\begin{split}
T_{eff,A1}(\nu) &= G_{a1}(\nu)G_{b1}(\nu)G_{c1}(\nu)(T_{in} + 
T_{eq,a1}(\nu))\\
&+G_{b1}(\nu)G_{c1}(\nu)T_{eq,b1}(\nu)\\
&+G_{c1}(\nu)T_{eq,c1}(\nu)~.
\end{split}
\end{equation}
The indices $a1, b1$ and $c1$ relate the respective gains and 
equivalent noise temperatures to the beamsplitter, to 
window 1 and the IR filter 1. For each element, the emissivity 
is written in terms of an equivalent input noise 
temperature which is given by the formula
\begin{equation}
\label{eq:02}
\begin{split}
T_{eq}(\nu) = T_{phys} \frac{1 - G(\nu)}{G(\nu)}~,
\end{split}
\end{equation}
with $T_{phys}$ being the physical (ambient) temperature of the 
element. Similarly, the effective load temperature $T_{eff,A2}(\nu)$ 
is calculated where only the dewar window and
the IR filter contribute.
Figures~\ref{fig:02}(a) and 
\ref{fig:02}(b) show the results for the effective temperatures. 
Figures~\ref{fig:02}(c) and \ref{fig:02}(d) present
equivalent circuits for the window-IR filter-beamsplitter and
window-IR filter cascade in front of mixer input port A1 
(LO port) and A2 (signal port).
\subsection{\label{sec2C}LO noise measurement 
using a balanced mixer}
\begin{figure}[tb]
\includegraphics[width=0.8\columnwidth]{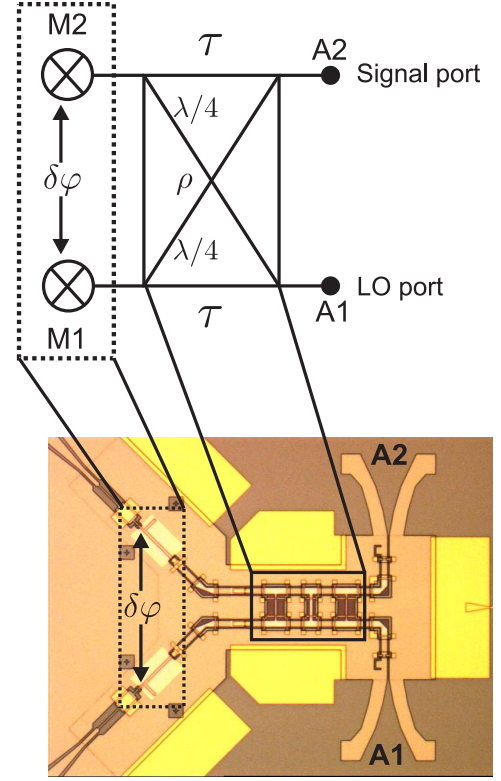}
\caption{\label{fig:03}Top: Balanced mixer circuit diagram 
showing the two mixer input ports A1 and A2, the 
$90^{\circ}$ hybrid coupler (crossed rectangle) 
and the two SIS mixers M1 and M2 having gains $G_{M1}$ 
and $G_{M2}$. $\tau^2$ and $\rho^2$ are power coupling 
factors and $\delta \varphi$ is the phase error of 
a possibly not ideal $90^{\circ}$ hybrid coupler.
Bottom: Detail photograph of the mixer chip focussing
on the RF part of the circuit.}
\end{figure}

A LO signal with sideband noise
is written in terms of a 
time-varying voltage
\begin{equation}
\label{eq:03}
V_{LO}(t) = \widetilde{V}_{LO} e^{i\omega_{LO}t} + 
V_{n}(t)~,
\end{equation}
where $\widetilde{V}_{LO}$ is the LO amplitude and 
$\nu_{LO} = \omega_{LO}/2\pi$ is the LO fundamental 
frequency. In Eq.~(\ref{eq:03}) the second term describes the 
near-carrier AM noise contribution to the LO signal
\begin{equation}
\label{eq:04}
V_{n}(t) = \sum_{k = -\infty}^{\infty} c_k e^{i\omega_k t}~.
\end{equation}
Generally, it has many frequency components 
$\omega_k$ near the carrier $\omega_{LO}$ and 
the expansion coefficients $c_k$ can in principle 
be determined by Fourier analyzing the signal.
In our experiment a blackbody signal with frequencies 
$\nu_S$ is applied to A2 
(signal port) and the LO signal together with noise (Eq.~\ref{eq:03}) 
is applied to A1 (LO port) of the probably not ideal balanced mixer shown in 
Fig.~\ref{fig:03}. A not ideal
balanced mixer has two mixers M1 and M2 with different gains 
$G_{M1} \not= G_{M2}$, the $90^{\circ}$ hybrid coupler is not symmetric with 
respect to the two output ports connecting the
two mixers, $\tau^2 \not= \rho^2$, and a phase error between the two output ports
of the $90^{\circ}$ hybrid coupler occurs, i.e.~$\delta \varphi \not= 0^{\circ}$.
This modifies the $Y$-factor of the receiver with respect to input A2 of the mixer 
which generally reads
\begin{equation}
\label{eq:05}
Y = \frac{T_{eff,A2,h} + T_{eff,A1,c}\cdot \frac{1}{NR} + \widetilde{T}_{rec}}
{T_{eff,A2,c} + T_{eff,A1,c}\cdot \frac{1}{NR} + \widetilde{T}_{rec}}
\end{equation}
for a non-ideal device, where $NR$ is the noise rejection ratio defined
by Eq.~(\ref{eq:06}). Equation~(\ref{eq:05}) is obtained by applying
the standard balanced mixer theory to our device in which $NR$ contains all
imperfections of the mixer. 

At the $\Sigma$ output of a $180^{\circ}$ IF hybrid coupler shown in 
Fig.~\ref{fig:01}, the combined (sum) current from the SIS mixers M1 and 
M2 can be measured. At the $\Delta$ output of the same IF hybrid coupler the IF current
of M2 is shifted by an additional phase of $-\pi$ with respect to the IF current of
M1 and the resulting IF (difference) current can be measured.
Reversing the bias polarity for M1 or M2, 
the output ports of the $180^\circ$ IF hybrid coupler
are exchanged, i.e.~$\Delta \leftrightarrow 
\Sigma$. For example, reading out the signals at the 
$\Sigma$ output of the IF hybrid coupler and sweeping the 
bias voltage of M1 from negative to positive values while 
M2 is kept constant at a negative voltage, shows the 
noise power from the LO port in the negative 
half of the voltage sweep and the signal power in the 
positive half without having to swap the power meter 
between the two IF hybrid output ports.

A more detailed discussion of the balanced mixer theory can be found in
\cite{maas1986,kooi2004,kooi2012}. The fundamental calculation of
the IF current of a SIS mixer or its gain is provided by Tucker and Feldman
in the framework of the quantum theory of mixing\cite{tucker1985} and is outside
of the scope of this paper.
\begin{figure*}[t]
\centering
\includegraphics[width=\textwidth]{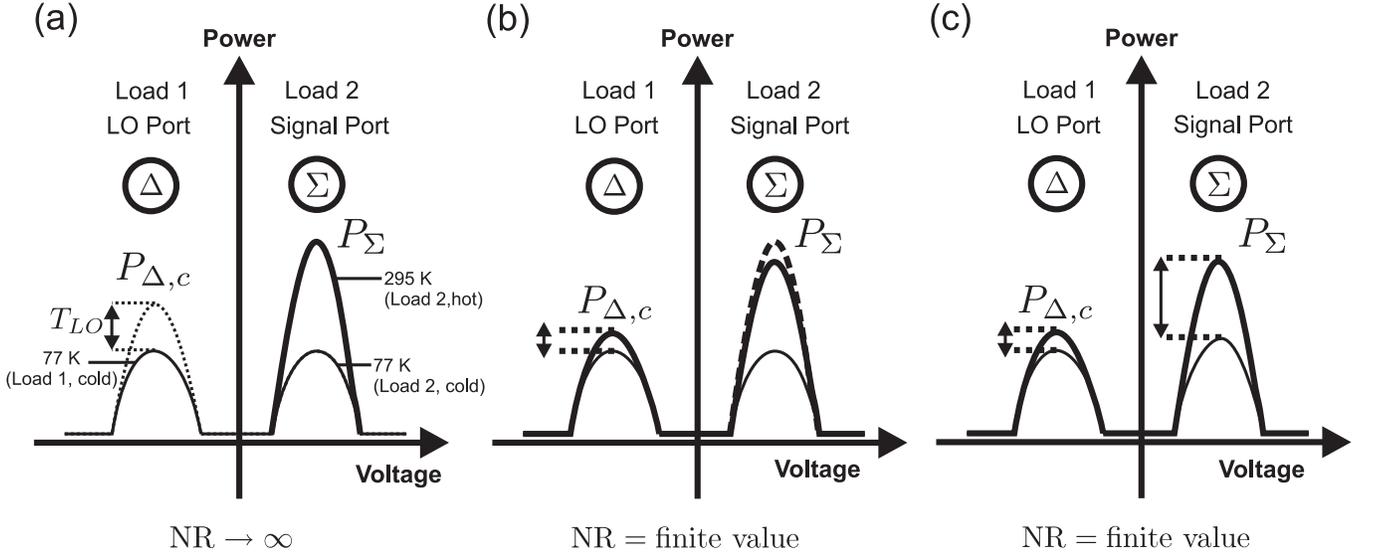}
\caption{\label{fig:04}Schematic representation of the 
balanced SIS mixer IF output power trace for three cases (a)-(c). 
One of the two SIS mixers on the balanced mixer chip is 
constantly biased with a negative voltage within the 
first photon-assisted tunneling step. A voltage sweep is 
applied to the other mixer. 
Power coming from load 1 (LO port) is measured 
in the $\Delta$ output ($P_{\Delta,c}$) whereas power coming 
from load 2 (signal port) is measured in the $\Sigma$ output 
($P_{\Sigma}$) of the IF output power trace. Thick solid 
lines show measurements where load 2 has a temperature of 
295~K whereas thin solid lines show measurements with a load temperature 
of 77~K. Load 1 has a constant temperature of 77~K. 
For each case it is indicated whether the noise 
rejection value ($NR$, Eq.~(\ref{eq:06})) is infinite or 
takes a finite value. (a) Ideal balanced 
mixer. Temperatures are equivalent values through the relation 
$P= T k_B B G_{rec}$, where $G_{rec}$ is the receiver gain, $T$ is the 
input temperature of either 295 or 77~K and $B$ is the IF bandwidth. 
The effect of near-carrier LO noise with equivalent noise temperature $T_{LO}$ 
is to increase the power $P_{\Delta,c}$ above the corresponding 
input noise power $77\cdot k_B B$ from load 1. (b) shows the effect of unequal 
mixer gain on the balanced mixer's IF output power 
for $\rho^2 = \tau^2 = 1/2$ and $\delta \varphi = 0^{\circ}$ or the effect of an asymmetry in the 
$90^{\circ}$ hybrid coupler for equal mixer gain and $\delta \varphi = 0^{\circ}$.
The dashed line indicates the ideal case for the $\Sigma$ output in which the 
distance between the dotted lines in the $\Delta$ output 
approaches zero. 
(c) Most likely situation during 
the experiment. In order to achieve the best balanced mixer 
performance, the difference between the two traces 
in the $\Delta$ output has to be minimized while 
maximizing the difference of the two traces in the 
$\Sigma$ output (arrows) which is achieved by adjusting the
phase shifter in the IF path (Fig.~\ref{fig:01}).}
\end{figure*}

In summary, in this paper the negative half of the voltage sweep 
represents the window 1 (LO noise) port and the positive half 
represents the window 2 (signal) port. This provides the 
possibility to determine $T_{rec}$ and near-carrier LO noise 
in one voltage sweep of mixer M1 or M2 from e.g.~-4~mV to 
+4~mV while the other mixer bias voltage is kept 
constant within the first photon-assisted tunneling 
step (Fig.~\ref{fig:08}(b)). The combined signal is 
read-out at the $\Sigma$ output of the IF hybrid coupler, 
see Figs.~\ref{fig:04} and \ref{fig:05}.

For a better understanding of our measurement results, a qualitative 
comparision of an ideal gain performance of the mixers and 
the $90^{\circ}$ hybrid coupler with a non-ideal device
is discussed below by using Fig.~\ref{fig:04}.
For simplicity, we assume for the moment that windows 1 
and 2, both IR blocking filters and the beamsplitter 
have a transmissivity of 1, i.e.~the temperature of load 1 
and 2 is not modified by the dielectrics. Furthermore, 
for our qualitative explanation in this section we 
assume at first a noiseless LO and then describe how the balanced 
mixer response changes when the LO has near-carrier noise.
\begin{description}
\item[Case 1] $G_{M1}=G_{M2}$, $\rho^2 = \tau^2 = 1/2$ and
$\delta \varphi = 0^{\circ}$.\\
This is the ideal case and
is qualitatively shown in Fig.~\ref{fig:04}(a). 
If the LO would have near-carrier noise of equivalent 
temperature $T_{LO}$, a larger noise power than 
just the noise power from a 77~K load is measured.
The same effect would occur for the following cases 2-3.
In order to better understand the non-ideal balanced mixer 
response, for clarity these cases are discussed assuming no LO noise.
\item[Case 2] $G_{M1}\not=G_{M2}$, $\rho^2 = \tau^2 = 1/2$ and
$\delta \varphi = 0^{\circ}$ or\\
$G_{M1}=G_{M2}$, $\rho^2 \not= \tau^2 = 1/2$ and
$\delta \varphi = 0^{\circ}$.\\
This case is illustrated in Fig.~\ref{fig:04}(b) for a 295~K load (thick solid line)
in front of window 2 (signal port) and for a 77~K load in front of window 1 (LO port).
The measured noise power $P_{\Delta,c}$ is in excess
of $77\cdot k_B B$ (arrow) and the height of $P_{\Sigma}$ for a 
295~K load is lower than in the ideal case (dashed line).
When translating near-carrier LO 
noise power into an equivalent noise temperature, 
the difference between the measured noise power in 
the $\Delta$ output when load 2 (signal port) is varied between 
77~K and 295~K is a major uncertainty of 
the measurement. 
\item[Case 3] $G_{M1} \not= G_{M2}$, $\rho^2 \not= \tau^2$ and 
$\delta \varphi \not= 0^{\circ}$.\\
This is the most realistic case observed in the 
experiment and is shown in Fig.~\ref{fig:04}(c). 
Both the negative and the positive half of 
the IF output power trace are influenced from 
the measured noise power. 
During the experiment the minimum of $T_{rec}$ 
measured at the $\Sigma$ output and the minimum of 
residual noise power from load 2 measured at the $\Delta$ output, 
is adjusted with the phase shifter,
therefore, finding the maximal value for 
$NR$. 
\end{description}
\section{\label{sec3}LO Noise Measurements}
\subsection{\label{sec3A}Quantitative LO noise determination}
This section presents heterodyne measurements using the 
balanced SIS mixer with three different LO sources: a Gunn LO 
and a synthesizer LO which uses two different power pre-amplifiers 
(PA1 and PA2 in Fig.~\ref{fig:01}). 
We use the mixer as a noise meter and quantify 
residual near-carrier LO noise power produced 
by the LO. The synthesizer LOs are optimally operated, 
i.e.~with input power saturating all 
of its components. For the measurements the output
power of PA1 or PA2 was set to its maximum value 
and the synthesizer output power was
16~dBm. Adjustment of the optimal coupled power of the 
synthesizer LO to the mixer was achieved using a 
polarization grid (Fig.~\ref{fig:01}).
The Gunn driven LO was used for comparison measurements as we
expect that this device produces less near-carrier LO noise than the
synthesizer driven LO.

\begin{figure*}[tb]
\includegraphics[width=0.6\textwidth]{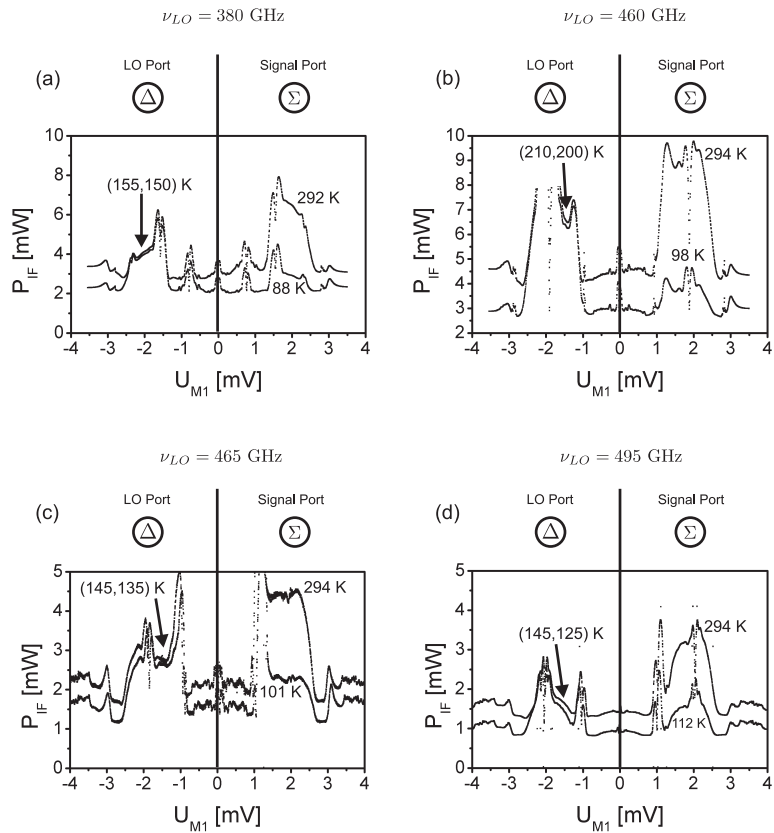}
\caption{\label{fig:05}Total IF output power of the
balanced SIS mixer as a function of $U_{M1}$ 
measured over the IF frequency range
4-8~GHz with a power meter. Mixer M2 is constantly 
biased at a negative voltage within the first 
photon-assisted tunneling step resulting in 
a relative shift between a pair of measurements for 
voltages larger or smaller than the photon-assisted 
tunneling step of M1 (for example from -1~mV to 1~mV).
This detail is omitted in Fig.~\ref{fig:04}. 
Temperature values in the $\Sigma$ output of the 
figure indicate the effective temperature of load 2 as 
seen from mixer input port A2 (signal port).  
The $\Delta$ output measures the total temperature 
as seen from mixer input port A1 (LO port), 
i.e.~the effective temperature of load 1 together with 
a possible temperature contribution from
the LO caused by near-carrier noise power. Arrows point 
to the bias region in which the temperature is measured. 
The traces in (a) and (b) were measured using the 
synthesizer driven LO whereas the traces in
(c) and (d) were measured using the Gunn driven LO.}
\end{figure*}
We start by describing our method of measuring near-carrier
LO noise and how this method is implemented in our experiment. 
One of the two mixers M1 or M2 is kept at a constant bias 
voltage within the first photon-assisted tunneling step 
of the SIS IV curve. The total IF output power is measured 
at the $\Sigma$ output of the $180^{\circ}$ IF hybrid 
coupler for opposite bias polarity of M1 and M2 and 
for two different temperatures $T_{L2}$ (295~K and 77~K) of 
load 2 while the $\Delta$ output is terminated with a 
$50~\Omega$ load. During all measurements load 1 is kept at 
a constant temperature of $T_{L1} = 77$~K.
For opposite bias polarity of M1 and M2, we denote 
the total IF output power by $P_{\Sigma}$, measured at the 
$\Sigma$ output. This signal is the one shown in
in the positive half 
of the power traces in Fig.~\ref{fig:05}. 
An analogous measurement is performed with same bias
polarity of the two mixers where we measure $P_{\Delta}$
at the $\Sigma$ output. This signal is the one shown in
the negative half of the power traces.

Figure~\ref{fig:05} shows typical results of 
the measured balanced mixer total IF output power 
using the synthesizer LO [(a) and (b)]
and the Gunn LO [(c) and (d)].
We measure the noise power emitted from load 2 
received by mixer input port A2 (signal port) and the 
superposition of noise power emitted from load 1 with 
near-carrier LO noise power which is received by mixer input port 
A1 (LO port). The combined signals are read-out at the 
$\Sigma$ output of the $180^\circ$ IF hybrid coupler shown 
in Fig.~\ref{fig:01}.

An ideal balanced mixer separately measures $P_{\Sigma}$ and
$P_{\Delta}$ and provides the possibility to quantify 
any noise signal received by mixer input port A1 (LO port), 
contained in $P_{\Delta}$ and shown in the negative part of the IF
output power trace, independent of the signal $P_{\Sigma}$ 
received by A2 (signal port) and which is shown in the positive part
of the IF output power trace. The isolation between the two 
distinct signals $P_{\Sigma}$ and $P_{\Delta}$ determines the quality 
of the balanced mixer. A criterion for the isolation 
is the noise rejection value $NR$ defined as
\begin{equation}
\label{eq:06}
NR = \frac{dP_{\Sigma}}
{dP_{\Delta}}~.
\end{equation}
In the equation, $dP_{\Sigma}$ is the 
IF output power difference for opposite mixer
bias polarity when load 2 has a temperature of 
295~K and 77~K and $dP_{\Delta}$ is the IF 
output power difference for same bias polarity.

For the rest of the paper we assume that $NR$ is sufficiently large so that
Eq.~(\ref{eq:05}) takes the form of Eq.~(\ref{eq:08}). The values for $NR$ 
which we determine from our measurements, summarized in Table~\ref{tab:02}, 
justifies this simplification. For the lowest $NR$ values, the introduced uncertainty
to the measured value of $\widetilde{T}_{rec}$ is approximately one standard deviation and
for higher $NR$ values, accordingly, it is less.

For the data analysis, in the negative and positive half 
of the IF output power trace the noise power from 
the two loads and the near-carrier LO noise power 
can be quantified (compare with Fig.~\ref{fig:04}). 
The Gunn LO is tuned to frequencies of 445, 465 and 
495~GHz. Measurements at frequencies 380, 420, 460 and 
490~GHz use the synthesizer LO.

\begin{table*}[htb]
\caption{\label{tab:02}Summary of the LO noise 
measurements using a balanced SIS mixer. The column 
summarizing the values for $T_{eff,A1,c}$ includes a second 
number in brackets which is the value $T_{eff,A1,c} + T_{LO}$. 
The equivalent LO noise temperature, $T_{LO}$, is listed 
in a separate column. At 465~GHz, the Gunn driven LO has very little 
output power resulting in an unusal large $\widetilde{T}_{rec}$ value
compared to the result using a similar frequency for the 
synthesizer driven LO.}
\begin{ruledtabular}
\begin{tabular}{ccc|l|l|l|l|l}
$T_{L1}$~[K]&$T_{L2}$~[K]
&$\nu_{LO}$~[GHz]&
$T_{eff,A1,c}$~[K]&$T_{eff,A2}$~[K]&
$\widetilde{T}_{rec}$~[K]&$T_{LO}$~[K]&NR~[dB]\\
\hline
\multicolumn{8}{c}{Gunn LO}\\
\hline
\multirow{2}{*}{77}&295&
\multirow{2}{*}{445}&$122~(145)\pm 2$&$294 \pm 2$&
\multirow{2}{*}{$90 \pm 5$}&
\multirow{2}{*}{$15 \pm 10$}&
\multirow{2}{*}{$12 \pm 1$}\\
&77&&122~(128)&91&&&\\
\hline
\multirow{2}{*}{77}&295&
\multirow{2}{*}{465}&108~(145)&294&
\multirow{2}{*}{100}&
\multirow{2}{*}{$32 \pm 6$}&
\multirow{2}{*}{$14$}\\
&77&&108~(135)&101&&&\\
\hline
\multirow{2}{*}{77}&295&
\multirow{2}{*}{495}&110~(145)&294&
\multirow{2}{*}{50}&
\multirow{2}{*}{$15 \pm 10$}&
\multirow{2}{*}{$10$}\\
&77&&110~(125)&112&&&\\
\hline
\multicolumn{8}{c}{Synthesizer LO}\\
\hline
\multirow{2}{*}{77}&295&
\multirow{2}{*}{380}&123~(155)&292&
\multirow{2}{*}{80}&
\multirow{2}{*}{$30 \pm 4$}&
\multirow{2}{*}{$15$}\\
&77&&123~(150)&88&&&\\
\hline
\multirow{2}{*}{77}&295&
\multirow{2}{*}{420}&117~(165)&294&
\multirow{2}{*}{90}&
\multirow{2}{*}{$38 \pm 10$}&
\multirow{2}{*}{$10$}\\
&77&&117~(145)&82&&&\\
\hline
\multirow{2}{*}{77}&295&
\multirow{2}{*}{460}&109~(210)&294&
\multirow{2}{*}{40}&
\multirow{2}{*}{$96 \pm 6$}&
\multirow{2}{*}{$15$}\\
&77&&109~(200)&98&&&\\
\hline
\multirow{2}{*}{77}&295&
\multirow{2}{*}{490}&109~(170)&294&
\multirow{2}{*}{60}&
\multirow{2}{*}{$56 \pm 6$}&
\multirow{2}{*}{$12$}\\
&77&&109~(160)&111&&&
\end{tabular}
\end{ruledtabular}
\end{table*}
The IF output power values $P_{\Sigma}$ for the
individual LO frequencies obtained for the 
temperatures $T_{L2}$ of load 2 
have to be related to the respective effective 
temperatures as seen by the two mixer input ports 
A1 and A2. For this purpose, a 
$Y$-factor measurement determines $T_{rec}$.
Furthermore, the $Y$-factor relates 
$P_{\Sigma}$, $T_{L2}$ and $T_{rec}$
\begin{equation}
\label{eq:07}
Y = \frac{T_{L2,h} + T_{rec}}{T_{L2,c} + T_{rec}} 
= \frac{P_{\Sigma,h}}{P_{\Sigma,c}}~,
\end{equation}
where $P_{\Sigma,h}$ and $P_{\Sigma,c}$ are the receiver 
IF output powers measured when load 2 (signal port) has a 
temperature of $T_{L2,h} = 295$~K and $T_{L2,c} = 77$~K.
Equation~(\ref{eq:07}) yields the same result, but is more 
suitable for our experiment when correcting
$T_{L2}$ and $T_{rec}$ for the thermal noise contribution 
coming from the dielectric slabs in front of mixer input port A2
\begin{equation}
\label{eq:08}
Y = \frac{T_{eff,A2,h} + \widetilde{T}_{rec}}
{T_{eff,A2,c} + \widetilde{T}_{rec}} 
= \frac{P_{\Sigma,h}}{P_{\Sigma,c}}~.
\end{equation}
In the equation above, $T_{eff,A2,h}$ is the effective
temperature seen by mixer input port A2 when load 2 
(signal port) has a temperature of 295~K ("hot load", h) 
while $T_{eff,A2,c}$ is
the effective temperature for a load temperature of 
77~K ("cold load", c). $\widetilde{T}_{rec}$ is the corrected 
receiver noise temperature with respect to the effective temperature.
Table~\ref{tab:02} summarizes the measured values for
$\widetilde{T}_{rec}$ showing the close to quantum limited 
performance of our receiver which is in the range of 
2-4 $h\nu_S/k_B$. In the same table, the column 
summarizing the effective temperatures $T_{eff,A1,c}$ 
seen by mixer input port A1 (LO port) contains two numbers. 
The first number outside of 
the brackets is the effective temperature calculated in 
Sec.~\ref{sec2B} for the temperature 77~K of load 1
and is used for comparision. The second number inside 
of the brackets is the effective temperature plus the
equivalent noise temperature of the LO
obtained from a calculation using the equation
\begin{equation}
\label{eq:09}
P_{\Delta,c} = \frac{T_{eff,A1,c} + T_{LO} + \widetilde{T}_{rec}}
{T_{eff,A2,h} + \widetilde{T}_{rec}}P_{\Sigma,h}
\end{equation}

\begin{figure}[tb]
\includegraphics[width=\columnwidth]{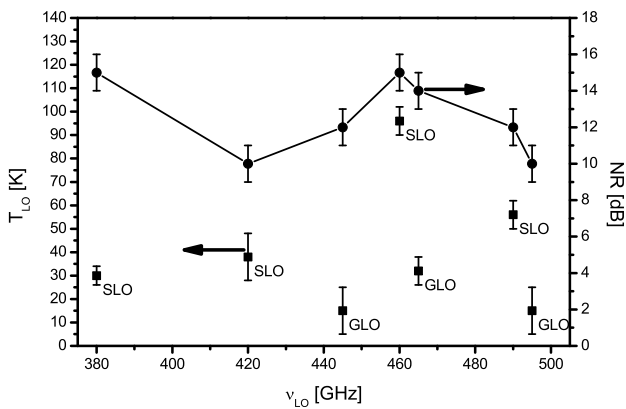}
\caption{\label{fig:06}Equivalent noise temperature $T_{LO}$ 
({\tiny$\blacksquare$}) and noise rejection $NR$ (\textbullet) as a function of LO 
frequency $\nu_{LO}$. Datapoints labeled with the abbreviation 
SLO were measured using the synthesizer LO whereas datapoints 
labeled with GLO belong to measurements which used the Gunn LO.}
\end{figure}
In the following we explain in detail the idea behind 
this equation and its derivation. The starting point 
is Eq.~(\ref{eq:08}). 
A perfect balanced mixer is symmetric with respect to its
two output ports and assuming for the moment that no additional 
external noise source (such as the noise power coming from the LO in 
our experiment) is present this allows to conduct a $Y$-factor measurement 
with either load 1 (LO port) or load 2 (signal port) with the same 
measurement result.
To measure the IF output power, the mixers M1 and M2 would have same bias
polarity when load 1 is used as a calibration blackbody source
and opposite bias polarity when load 2 is used.
It would even be possible to use the two loads together, say 
with load 1 having a temperature of 77~K and load 2 having a 
temperature of 295~K, again using the $\Sigma$ output of the 
IF hybrid coupler. Using Eq.~(\ref{eq:08}), with
$P_{\Sigma,h}$ being the measured noise power from load 2 and 
$P_{\Sigma,c} = P_{\Delta,c}$ being the measured noise power from 
load 1, would give the same $Y$-factor as before using only one 
load for the measurement. If an unknown amount of noise 
power is superimposed with the noise power coming 
from one of the two loads, the $Y$-factor changes and gives us the opportunity to 
measure the noise power. In our experiment the noise power 
coming from load 1 (LO port) having a temperature of 77~K is increased 
by the near-carrier LO noise power (Fig.~\ref{fig:04}(a)). 
We measure first the $Y$-factor, $Y = P_{\Sigma,h}/P_{\Sigma,c}$, which we
obtain from a measurement using load 2 (signal port), having two different
temperatures of $T_{L2,h}=295$~K and $T_{L2,c}=77$~K. 
From this measurement, $\widetilde{T}_{rec}$
is calculated using Eq.~(\ref{eq:08}) and is 
summarized in Table~\ref{tab:02}. The indicated 
uncertainty is the root-mean-square deviation which we obtained
from a series of identical measurements.
In an ideal balanced mixer, noise power received by mixer input 
port A1 (LO port) can be independently 
measured from the noise power received by mixer input port 
A2 (signal port), therefore, we substitute 
$P_{\Sigma,c} \rightarrow P_{\Delta,c}$. 
In order to fulfill Eq.~(\ref{eq:08}), we substitute 
$T_{eff,A2,c} \rightarrow T_{eff,A1,c} + T_{LO}$.
Solving for $P_{\Delta,c}$ results in Eq.~(\ref{eq:09}).
$T_{LO}$ is used as a free parameter in order to fit 
the right-hand side of this equation to the measured value
of $P_{\Delta,c}$. The two numbers enclosed by the brackets in 
Table~\ref{tab:02} are the values $T_{eff,A1,c} + T_{LO}$, 
obtained for load 2 temperatures $T_{L2,h}=77$~K and
$T_{L2,h}=295$~K. The mean value of the two measurements
is summarized in the $T_{LO}$ column for each LO frequency
investigated in our experiment. A conservative estimate for the
uncertainty in $P_{\Sigma}$ and $P_{\Delta}$ is
0.1~mW (1\% - 3\%) due to the power variation 
within the bias region where the noise power is measured. 
The uncertainty in $T_{eff,A1,c}$ and
$T_{eff,A2}$ is determined from the fit to the measured 
transmissivity using the FTS described in 
Sec.~\ref{sec2B} and the result is 2~K. 
The uncertainty of $T_{LO}$ is thus determined by the 
uncertainty of the fit to the FTS transmissivity data and by the
difference between the mean value of $T_{eff,A1,c} + T_{LO}$ 
and the single measurements when load 2 has temperatures of either
$T_{L2} = 295$~K or 77~K. This difference is a result of non-ideal
balanced mixer performance. The noise rejection $NR$ is corrected
for the transmissivity of the dielectric slabs in front 
of mixer input port A2 and its uncertainty is mainly dominated 
by the uncertainties of $P_{\Sigma}$ and $P_{\Delta}$.
All results are summarized in Table~\ref{tab:02} and 
in Fig.~\ref{fig:06}.

Measurements with the 
synthesizer driven LO for the two frequencies 380~GHz and 420~GHz 
used the power pre-amplifier PA1 and for the two higher frequencies
460~GHz and 490~GHz the power pre-amplifier PA2. Evidently, within
the tolerance of our measurement, using PA2 
an up to three times larger LO noise contribution 
could be measured compared to measurements using PA1. 
A measurement using the Gunn driven LO at 465~GHz resulted in the
highest value of $T_{LO}$ for this LO device, however, which was
comparable with the synthesizer driven LO measurements at 380~GHz and
420~GHz. Gunn driven LO measurements at 445~GHz and 495~GHz resulted
in the lowest values for $T_{LO}$. These values are a factor of 2
smaller than the LO noise
values obtained with the synthesizer driven LO using PA1 and a factor of
4 - 6 smaller than the values obtained with the synthesizer driven LO
using PA2.
\subsection{\label{sec3B}Effect of LO noise on single ended 
mixer performance}
This section investigates the impact of LO 
noise power on a single-ended mixer which does not 
provide noise rejection like a balanced mixer. The results 
are used to verify our results from the previous section. 
In particular, our findings are interesting
for a better understanding of the noise performance of 
single-ended mixers working in a frequency range 
where no or only few balanced mixer technologies are 
available.

\begin{figure}[t]
\includegraphics[width=0.60\columnwidth]{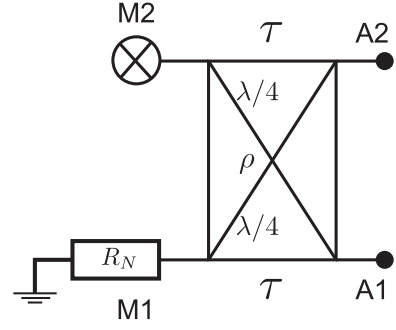}
\caption{\label{fig:07}Circuit diagram showing the
measurement principle used to study the impact of 
LO noise power on a single-ended mixer. For an ideal 
$90^{\circ}$ hybrid coupler, half of the signals 
received by the mixer input ports A1 and A2 are 
detected by M2 whereas half of the signals are absorbed 
by mixer M1. This is the case for $\tau^2 = \rho^2 = 1/2$.
M1 is operated in the normal conducting 
state with resistance $R_N\approx 21~\Omega$.
Because of symmetry, the same principle applies 
to M2 being operated in the normal conducting 
state where now M1 is used as mixer.}
\end{figure}

Figure~\ref{fig:07} shows a circuit diagram illustrating our
measurement method employed to determine the influence of LO
noise power on the single-ended mixer performance. The balanced mixer
circuit, used in the previous section, is operated with 
one of two mixers voltage biased far above the
superconducting gap voltage, therefore, being in the normal 
conducting state. This mixer is used as an absorber, terminating
one branch of the $90^{\circ}$ hybrid coupler. It is important 
to mention that in our device, noise generated by a 
mixer in the normal conducting state does not influence 
the heterodyne measurement using the other mixer by adding 
e.g.~additional noise. Following the quantum tunneling theory, 
quasiparticles driven by a constant voltage $V$ 
through an SIS junction generate 
current noise which can be written as\cite{rogovin1974}
\begin{equation}
\label{eq:10}
\begin{split}
\left\langle i^2(\nu)\right\rangle  &= e 
\left[I\left(V + \frac{h\nu}{e}\right)
\coth\left(\frac{eV + h\nu}{2k_B T}\right)\right.
\\
&\left.+
I\left(V - \frac{h\nu}{e}\right)
\coth\left(\frac{eV - h\nu}{2k_B T}\right)
\right]~,
\end{split}
\end{equation}
illustrating a fluctuation-dissipation relation which connects
the spectral quasiparticle current noise per unit bandwidth 
to the quasiparticle current 
response $I(\ldots)$.
For large voltages $eV\gg k_BT$ and low frequencies for 
which $eV\gg h\nu$, the current noise is 
$\left\langle i^2(\nu)\right\rangle \approx 2eI(V)\coth(eV/2k_BT)$
which is the usual shot noise formula, 
to good approximation independent of $\nu$.
Intermediate frequency shot noise is not transmitted by the MIM capacitors 
separating the two SIS junction circuits in the balanced 
mixer\cite{westig2011} until $\nu = 30$~GHz. 
For high frequencies the current noise approaches thermal equilibrium
\begin{equation}
\label{eq:11}
\left\langle i^2(\nu)\right\rangle \approx e\left[I\left(\frac{h\nu}{e}\right) - 
I\left(-\frac{h\nu}{e}\right)\right]\coth\left(\frac{h\nu}{2k_BT}\right)~.
\end{equation}
In this limit the noise arriving 
at the other SIS junction circuit on the balanced mixer 
chip is determined by the isolation of the two branches of 
the $90^{\circ}$ hybrid coupler. In the detection bandwidth 
380-520~GHz, we derive by electromagnetic field simulation 
of the mixer chip an isolation value between the two mixers 
of -18~dB to -25~dB. This suggests that the contribution 
described by Eq.~(\ref{eq:11}) can be neglected as well.

\begin{figure}[tb]
\includegraphics[width=\columnwidth]{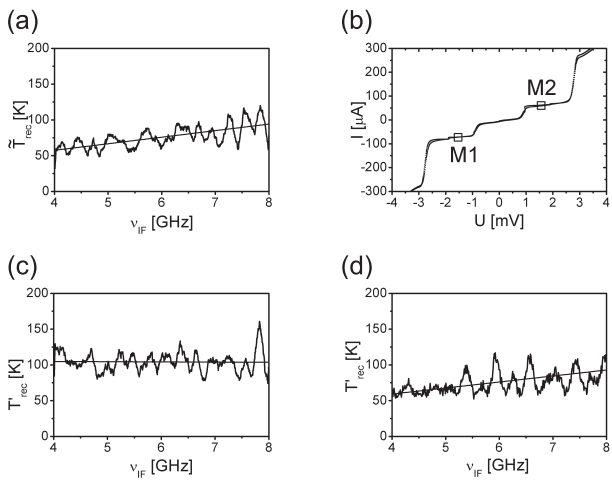}
\caption{\label{fig:08}
(a) $\widetilde{T}_{rec}$ of the balanced mixer for
$\nu_{LO} = 460$~GHz using a synthesizer LO.
(b) Rectangles show the bias voltages of the two 
mixers M1 and M2 during the measurement of the 
trace shown in (a). The figure shows two LO pumped 
SIS IV curves which are almost equal demonstrating 
the optimal device performance for this frequency.
In (c) a single-ended measurement with mixer M2 at 
$\nu_{LO} = 460$~GHz is presented which uses the synthesizer LO and 
during which $U_{M1} = -5$~mV. The receiver 
noise temperature is corrected for the effective load 
temperatures with respect to both mixer input ports 
A1 and A2. (d) shows the result of a single-ended measurement
using mixer M2 where again $U_{M1} = -5$~mV and
for $\nu_{LO} = 455$~GHz. 
Instead of using the synthesizer driven LO we 
employed a Gunn driven LO for the measurement 
which results in a similar performance
like observed in (a).}
\end{figure}

The other mixer is operated at voltages within 
the first photon-assisted tunneling step and is used as 
a heterodyne mixer.
An ideal $90^{\circ}$ hybrid coupler distributes
one half of each signal received by mixer input ports A1 and A2
to the two mixers. As can be seen in Fig.~\ref{fig:08}(b),
to a good approximation
this is the case for the frequency $\nu_{LO}=460$~GHz. 
The measurement of the single-ended mixer receiver noise temperature
for $\nu_{LO} = 460$~GHz with the synthesizer driven LO 
results in $T'_{rec} = 100$~K for M1 and
$T'_{rec} = 78$~K for M2. A standard $Y$-factor method was conducted 
where first the combination of the noise power of load 1 and 2 
both at a temperature of 77~K are measured using mixer M1 and M2 and 
subsequently the same measurement is
performed for both loads at a temperature of 295~K.
$T'_{rec}$ is the receiver noise temperature
without taking into account the noise power contribution of the LO
but after correcting for the noise contribution of the dielectric 
slabs in front of the mixer input ports A1 and A2.
In contrast to the result of $\widetilde{T}_{rec} = 40$~K 
obtained from our balanced measurement at 
$\nu_{LO} = 460$~GHz, by using only the single-ended mixer
obviously a larger receiver noise temperature
is measured. In the previous section, we 
found an LO noise contribution of $T_{LO} = 96$~K 
at 460~GHz.
This means, by assuming an ideal
coupler, the LO noise contributes
$96/2$~K noise to each of the two mixers 
in addition to the contribution of load 1 (LO port).
Considering the effective temperatures
$T_{eff,A1}$ and $T_{eff,A2}$ for a load temperature of 295~K 
and taking into account the additional LO noise power
contribution superimposed with the noise power emitted by load 1, 
results in a total effective temperature of 342~K referred to the 
input of M1 and M2. Similarly when load 1 and 2 have a temperature of 
77~K this results in a total effective temperature of 152~K with respect
to the input of both mixers. Recalculating the single-ended 
mixer receiver noise
temperature with these input temperatures results in
$\widetilde{T}_{rec} = 50$~K for M1 and 
$\widetilde{T}_{rec} = 30$~K for M2 reproducing the 
balanced receiver noise temperature within a range 
of 10~K and justifying our result for 
$T_{LO}$. The balanced receiver noise temperature is not 
altered by LO noise to good approximation, provided that the 
quality of the balanced mixer operation is 
sufficiently good, as in our experiment. 
Throughout all frequencies within the operation bandwidth 
of the mixer, we observe that the performance of M1 is slightly
inferior to that of M2 and we find that the balanced 
performance is about the average of both.
The expected effect of LO noise power on the performance of a 
single-ended mixer receiver is to add $T_{LO}$ to $T_{rec}$.
Using an LO with a significant noise power contribution 
and comparing $T_{rec}$ as a function of $\nu_{IF}$ for a 
balanced and a single-ended mixer an interesting difference between both IF 
traces is observed. In Fig.~\ref{fig:08} we show such a measurement
in (a) for the balanced mixer and in (c) for a single-ended measurement
using mixer M1. Both measurements were conducted with 
the synthesizer driven LO. From a linear fit
to the two IF traces we find for the balanced mixer measurement a positive
slope while for the single-ended mixer measurement 
we find a slightly negative slope. This effect is a strong evidence
for frequency dependent LO noise power. Our assumption is supported 
by an identical single-ended mixer measurement, however, using a 
Gunn driven LO instead of the synthesizer driven LO and which is shown in 
Fig.~\ref{fig:08}(d). The slope of the IF is clearly positive 
which suggests that less near-carrier noise power is produced 
by this LO compared to the result of Fig.~\ref{fig:08}(c). 
This is in agreement with the $T_{LO}$ value for the LO device 
used for the measurement shown 
in Fig.~\ref{fig:08}(d) which is 38~K and, therefore, in this measurement 
only 38/2~K additional noise is added to each of the two 
mixers on the balanced mixer chip. Moreover, we find 
that the measured single-ended noise performance using the 
Gunn driven LO is comparable to the balanced mixer 
measurement shown in Fig.~\ref{fig:08}(a).

The balanced mixer IF trace has a positive slope 
because the noise rejection has a certain bandwidth and 
decreases as a function of $\nu_{IF}$. For the measurement 
shown in Fig.~\ref{fig:08}(a) the balanced mixer is tuned 
to have the strongest noise rejection for the small IF frequencies whereas in 
Fig.~\ref{fig:08}(c) it is observed that LO noise power has 
the strongest influence for small values of $\nu_{IF}$. This leads to a 
negative slope of the IF trace which dominates the usually 
observed increase in the noise temperature 
for large $\nu_{IF}$ mainly due to the 
SIS junction capacitance (Fig.~\ref{fig:08}(d)).
Therefore, the effect of near-carrier noise power should 
be strongest for small IF frequencies.
\section{\label{sec4}Conclusion and Outlook}
To conclude, we have presented an experimental method which
allows to use a balanced mixer as a noise meter. 
We have shown that as a result of such a measurement the 
additional noise power contribution of an LO to an 
astronomical receiver system can be identified. 
Three LO sources were used in our heterodyne experiment which measured 
their equivalent noise temperature. The balanced mixer operation 
provides the measurement of $T_{rec}$ independent of
the noise contribution $T_{LO}$. The highest $T_{LO}$ value was 
measured for the frequency 460~GHz using the synthesizer driven 
LO. Components of this LO have low-frequency 
gain,\cite{privatecommhesler2012} so that for example 
even very small low-frequency amplitude noise 
contributions coming from our synthesizer can result at the
end in near-carrier LO noise power, measurable in the 
IF frequency range 4-8~GHz. An alternative
interesting explanation is given by Bryerton 
{\it et al.}\cite{bryerton2007} discussing the possibility of phase 
noise to amplitude noise conversion in a frequency multiplier chain.
If the frequency multiplier chain 
of the synthesizer driven LO does not significantly contribute to the 
noise power of the complete LO system\cite{erickson2004}, then 
the power amplifier is the most likely noise source of the LO chain.
The large measured difference in $T_{LO}$ when different power 
amplifiers were used (PA1 and PA2) indicates this.
In this case our measurements can be applied to a THz LO. 
Attaching the lower frequency active components of a THz LO 
to our lower frequency LO would provide a measurement method to 
characterize the THz LO noise performance, rendering precise 
characterizations of superconducting THz mixer devices possible.
In a heterodyne measurement where we
used only one of two mixers on the balanced mixer chip, 
our measurement shows the impact of LO noise power on 
single-ended mixers. Here the expected result is that 
$T_{LO}$ is added to $T_{rec}$. 
Measuring the receiver
noise temperature, the IF trace of a balanced mixer shows a positive
slope due to the bandwidth limited noise rejection. 
Near-carrier LO noise power effects the single-ended
IF frequency resolved receiver noise temperature in a different way
and the result is an increase of noise at lower frequencies, resulting
in an effectively flat or even slightly negative slope.
We expect that this effect should significantly
decrease the sensitivity of receivers employing HEB mixers, 
e.g.~for frequencies above 1.4~THz where these devices 
are the heterodyne mixers of choice and where only few 
balanced mixers were realized to date\cite{meledin2009,dochev2011}.
\begin{acknowledgments}
This work is carried out within the Collaborative Research 
Council 956, sub-project D3, funded by the Deutsche 
Forschungsgemeinschaft (DFG) and by BMBF, 
Verbundforschung Astronomie under contract no.~05A08PK2. 
The excellent support of the in-house
machine shop is gratefully acknowledged. 
M.~P.~Westig would like to express his 
gratitude to T.~W.~Crowe and  J.~Hesler from 
Virginia Diodes, Inc., USA, for fruitful discussions and
very useful hints concerning the synthesizer driven local 
oscillator and to J.~Childers from Spacek Labs Inc., USA,
for a very fast response on questions concerning the power 
amplifiers. All authors thank the Max-Planck-Institut 
f\"ur Radioastronomie in Bonn, Germany, for the possibility 
of using their 380-520~GHz synthesizer driven local oscillator source.
One of the authors (MPW) would like to thank the Bonn-Cologne
Graduate School of Physics and Astronomy for financial and 
travel support. The balanced mixer devices were fabricated
in the KOSMA microfabrication laboratory in K\"oln.
\end{acknowledgments}

\begin{thebibliography}{30}
\bibitem [{syn()}]{synthesizer}%
  \BibitemOpen
  \href@noop {} {}\bibinfo {note} {{Our synthesizer is an Agilent Technologies
  PSG CW signal generator, model number E8247C. The frequency range is 250~kHz
  - 20~GHz}}\BibitemShut {NoStop}%
\bibitem [{spe()}]{spectrumanalyzer}%
  \BibitemOpen
  \href@noop {} {}\bibinfo {note} {{The spectrum analyzer used throughout our
  measurements is of type PXA N9030A with a frequency bandwith of 3~Hz -
  26.5~GHz}}\BibitemShut {NoStop}%
\bibitem [{rat()}]{ratracecoupler}%
  \BibitemOpen
  \href@noop {} {}\bibinfo {note} {{We use a coupler from ET Industries, model
  number J-412-180. The frequency bandwith is 4.0 - 12.4~GHz}}\BibitemShut
  {NoStop}%
\end{thebibliography}
\end{document}